# Multifunctional Antiferromagnetic Materials with Giant Piezomagnetism and Noncollinear Spin Current

Hai-Yang Ma[1,†], Mengli Hu[2,†], Nana Li[2,†], Jianpeng Liu[2], Wang Yao[3], Jin-Feng Jia[1,4,*] and Junwei Liu[2,*]

[1] *Key Laboratory of Artificial Structures and Quantum Control (Ministry of Education), Shenyang National Laboratory for Materials Science, School of Physics and Astronomy, Shanghai Jiao Tong University, Shanghai 200240, China*

[2] *Department of Physics, The Hong Kong University of Science and Technology, Hong Kong, China*

[3] *Department of Physics and Center of Theoretical and Computational Physics, University of Hong Kong, Hong Kong, China.*

[4] *Tsung-Dao Lee Institute, Shanghai 200240, China*

*Correspondence to: jfjia@sjtu.edu.cn and liuj@ust.hk

†These authors contributed equally to this work.

**We propose a new type of spin-valley locking (SVL), named *C*-paired SVL, in antiferromagnetic systems, which directly connects the spin/valley space with the real space, and hence enables both static and dynamical controls of spin and valley to realize a multifunctional antiferromagnetic material. The new emergent quantum degree of freedom in the *C*-paired SVL is comprised of spin-polarized valleys related by a crystal symmetry instead of the time-reversal symmetry. Thus, both spin and valley can be accessed by simply breaking the corresponding crystal symmetry. Typically, one can use a strain field to induce a large net valley polarization/magnetization and use a charge current to generate a large noncollinear spin current. We predict the realization of the *C*-paired SVL in monolayer $V_2Se_2O$, which indeed exhibits giant piezomagnetism and can generate a large transverse spin current. Our findings provide unprecedented opportunities to integrate various controls of spin and valley with nonvolatile information storage in a single material, which is highly desirable for versatile fundamental research and device applications.**

Exploring new quantum degrees of freedom (DOFs) and finding new control approaches are crucial to realize multifunctional materials for developing the next-generation information technologies. For instance, the periodicity in a crystal can endow degenerate but inequivalent valleys in the electron dispersion, forming an emergent DOF that can encode information as spin[1,2]. Moreover, spin and valley DOFs can be further locked together to form the spin-valley locking (SVL) as proposed in transition metal dichalcogenides due to the strong spin-orbital coupling and the time-reversal symmetry[3]. SVL enables many dynamical control approaches[4] and dramatically increases the depolarization time for both spin and valley[5] since spin flipping needs to change valley index and vice versa[6–8]. However, it is very difficult to realize information storage (memory) in these SVL materials since the information is encoded by the spin and valley of carrier or current.

The discoveries of graphene and other 2D materials provide excellent material platforms to study these emergent quantum DOFs and their control approaches at the microscopic level [9–11]. Being atomically thin, 2D materials are easier to engineer via gate modulation[12,13], strain[14] and proximity effect[15]. Recent discoveries of atomic-thin ferromagnetic[16–20] and antiferromagnetic (AFM) materials[21–26] further enriched the 2D material family and opened a new era for spintronics and valleytronics with nonvolatile magnetic information storage. Moreover, 2D AFM materials possess many other desirable features such as higher-speed operation and lower-energy consumption as proposed in the AFM spintronics[27–30]. It has been demonstrated that the AFM order can be switched by an electric current as in CuMnAs-type AFM materials[31], while many other dynamic controls of spin and valley as in SVL materials have not yet realized. Currently, to simultaneously realize both the nonvolatile memory and dynamical controls of spin and valley, one usually needs to create heterostructures of SVL and 2D magnetic materials[32].

In this work, by coupling the valley DOF to the AFM order in a single material, we propose a new type of SVL, named $C$-paired SVL, which can simultaneously realize giant piezomagnetism and large noncollinear spin current, thus enabling both static and dynamical controls of spin and valley DOFs. We predict its materialization in monolayer $V_2Se_2O$ and other experimentally verified AFM materials using symmetry analysis and first-principles calculations.

In *C*-paired SVL, SOC is no longer a necessary condition and the valley-contrasted spin splitting can arise from strong exchange coupling of itinerant electrons to AFM order and hence can be as large as several eVs. Moreover, valleys forming the emergent quantum DOF here are related by a crystal symmetry instead of the time reversal symmetry, which makes both spin and valley DOFs much easier to access by versatile controls that lift the constraint from the corresponding symmetry. Typically, one can use strain to break the crystal symmetry and induce sizeable valley splitting, and hence generate a large valley polarization and a net magnetization upon a finite doping. Besides static approaches, one can use various dynamical approaches like a charge current to break the crystal symmetry easily and induce a large noncollinear spin current accordingly. Remarkably, a pure transverse spin current like in spin Hall effect[33] can be realized if the charge current is along the direction preserving the crystal symmetry. Compared with the conventional spin Hall effect originated from SOC, the transverse spin current here is from the anisotropy of Fermi pockets and the equivalent spin Hall angle could be much larger (~0.7 in monolayer $V_2Se_2O$). In addition, due to the weak or absence of SOC and protection of SVL as in conventional SVL systems, long spin depolarization time can be expected, which is critical for spintronics applications[34].

Compared to the conventional SVL and CuMnAs-type AFM materials[31], the *C*-paired SVL material can realize multiple functions in one single material and possesses advantages of different systems and many other unique properties. Typically, one can use both static AFM order and the spin/valley of carriers/current to encode and process information. The former one gives rise to non-volatile memory, and the latter one provides many other dynamic approaches as optical methods in conventional SVL materials[4]. Moreover, both static and dynamic information can be naturally transformed to each other in *C*-paired SVL materials since a spin current can affect the AFM order and vice versa. In addition, the magnetoelastic coupling intrinsically exists in *C*-paired SVL materials, which enables the direct conversion between magnetic and mechanical properties, realizing many unique phenomena like piezomagnetism and magnetostriction and making *C*-paired SVL materials very useful in various sensors[35].

Our findings provide a new class of multifunctional material platform that combines versatile static and dynamic controls of spin and valley with high-density nonvolatile information storage, and extend these studies into more general systems of no or weak SOC and collinear AFM systems, which opens up a variety of new opportunities for finding more suitable materials in device applications of spintronics and valleytronics.

**General theory of *C*-paired SVL and its unique properties**. Valley in a crystal refers to the degenerate energy extrema well separated in momentum space, and its strong coupling to the unfrozen spin realizes the SVL. As shown in Fig. 1a, for the conventional SVL in transition metal dichalcogenides, strong SOC gives rise to sizable spin splitting in the individual valley, where the time-reversal symmetry enforces the opposite sign of the splitting at the time reversal pair of valleys. This extensively explored form of SVL relies on both strong SOC and inversion symmetry breaking, and we refer it as *T*-paired SVL. Note that the electronic properties of *T*-paired SVL systems also depend on the crystal symmetries as bilayer TMDs[36,37] and strained $MoS_2$ [38].

We first show that one can also realize SVL in an AFM system even without SOC. As shown in Fig. 1b, the AFM order breaks the time-reversal symmetry and generally induces a spin splitting between states $\Psi_\mathbf{K}^\uparrow$ and $\Psi_\mathbf{K}^\downarrow$ through strong exchange couplings, i.e. $E^\uparrow \neq E^\downarrow$ if there is no *PT* symmetry[39]. One can prove that there must be another two spin-polarized states $\Psi_{\mathbf{K}'}^\uparrow$ and $\Psi_{\mathbf{K}'}^\downarrow$ but with switched energy $E^\downarrow$ and $E^\uparrow$ at the crystal symmetry related valley $\mathbf{K}' = C\mathbf{K}$, i.e. $H_{\mathbf{K}'}\Psi_{\mathbf{K}'}^\uparrow = E^\downarrow \Psi_{\mathbf{K}'}^\uparrow$ and $H_{\mathbf{K}'}\Psi_{\mathbf{K}'}^\downarrow = E^\uparrow \Psi_{\mathbf{K}'}^\downarrow$, where $C$ is the crystal symmetry connecting the two AFM sublattices (more details in Sec. IA of the Supplementary Information (SI)). In other words, a crystal symmetry instead of time-reversal symmetry can also lead to SVL, which we refer to *C*-paired SVL. It is worth emphasizing that the spin splitting here is from exchange couplings between itinerant electrons and local magnetic moments, thus it could be as large as several eVs and SOC is no longer necessary.

Different valleys in *C*-paired SVL form an emergent quantum DOF, which can be used to encode information as spin. Moreover, they are related by a crystal symmetry instead of the time-reversal symmetry as in *T*-paired SVL. Thus, one can use various approaches like a strain or electric field to break the crystal symmetry to access this quantum DOF easily, which will give rise to many interesting phenomena and applications like giant piezomagnetism and noncollinear spin current. In general, various crystal symmetries can lead to the *C*-paired SVL, and to make discussions concrete we will focus on the mirror symmetry in the following parts.

As shown in Fig. 1b, there are two valleys related by a mirror symmetry $M_\phi$, where $\phi$ is the angle between the mirror plane and the $x$ axis (defined as the main axis of the valley with spin-up polarization), and a strain can easily break the mirror symmetry to induce a valley polarization. Using the deformation potential theory[40], the leading effect of a strain on the band edge of a gapped system is to shift the energy by $\text{Tr}(D\varepsilon)$, where $\varepsilon$ is the strain tensor and $D$ is the deformation potential tensor. By using $D' = M_\phi D M_\phi^{-1}$, we can get the strain-induced energy difference between the two valleys as $\delta E = [(D_{xx} - D_{yy})\sin^2 2\phi - D_{xy}\sin 4\phi](\varepsilon_{xx} - \varepsilon_{yy}) + [4D_{xy}\cos^2 2\phi - (D_{xx} - D_{yy})\sin 4\phi]\varepsilon_{xy}$. Considering a uniaxial strain $\varepsilon_\theta$ along $\theta$ direction, the energy shift is $\delta E = [(D_{xx} - D_{yy})\sin 2\phi - 2D_{xy}\cos 2\phi]\sin(2\phi - 2\theta)\varepsilon_\theta$ (more details in Sec. IB of SI). Since different valleys have opposite spin polarization, upon finite doping with carrier density $n$, there will be a net static magnetization $M = \int_{-\infty}^{E_f(n)}[\rho^\uparrow(\varepsilon) - \rho^\downarrow(\varepsilon)]d\varepsilon$, where $E_f$ is the Fermi level and $\rho^{\uparrow,\downarrow}$ is the spin-up and spin-down density of states and can be assumed as a constant $\rho$ around the band edge of a 2D gapped system. The strain-induced magnetization is then obtained as

$$M(\theta, \varepsilon_\theta, n) = \begin{cases} -\gamma \sin(2\phi - 2\theta)\varepsilon_\theta, & 0 \le \varepsilon_\theta < \dfrac{n}{|\gamma \sin(2\phi - 2\theta)|} \\ -\text{sign}(\gamma \sin(2\phi - 2\theta))n, & \varepsilon_\theta \ge \dfrac{n}{|\gamma \sin(2\phi - 2\theta)|} \end{cases}$$

, where $\gamma = \rho[(D_{xx} - D_{yy})\sin 2\phi - 2D_{xy}\cos 2\phi]$ is a constant depending on the materials details. As shown in Fig. 1c, the magnetization linearly increases with strain until all the carriers are fully polarized at a fixed doping. In addition, the magnetization depends on the strain direction with a periodicity of $\pi$. Piezomagnetism is the strongest for $\theta = \phi \pm \pi/4$ and is zero for $\theta = \phi$ or $\phi \pm \pi/2$, which is consistent with the symmetry analysis because a normal strain in or perpendicular to the mirror plane does not break the mirror symmetry. Moreover, the magnetization here is not because of the realignment of local magnetic moments as in conventional piezomagnetism[41], but from strain-induced occupation imbalance between spin-up and spin-down electrons, which is mainly determined by electronic property anisotropy and could be very large to realize a giant piezomagnetism. Note that, *T*-paired SVL systems can also realize a net magnetization through breaking the time-reversal symmetry as using a charge current in strained $MoS_2$[38], while it is dynamic properties and disappears once the charge current is switched off. On the contrary, in *C*-paired SVL system, it is a static property due to piezomagnetism arisen from the intrinsic magnetoelastic coupling. This coupling can also lead to magnetostriction where one can induce a strain using an external magnetic field. All these properties make *C*-paired SVL materials good candidates for magnetic sensors and applications in other fields [42].

Besides a static strain field, a charge current under an electric field can also dynamically induce the imbalance between the two valleys in *C*-paired SVL, and spontaneously generate a noncollinear spin current from the anisotropy of valleys with different spins. Around the band edge of a gapped system, the dispersion can be generally approximated as $E^K(\boldsymbol{k}) = \frac{\hbar^2 k_x^2}{2m_1} + \frac{\hbar^2 k_y^2}{2m_2}$, where $m_1$ and $m_2$ are the effective mass. Based on the Boltzmann equation and mirror symmetry $M_\phi$, the spin conductivity under an electric field along $\theta$ direction (Fig. 1d) can be calculated as

$$\sigma^S = \sigma^K - \sigma^{K'} = \sigma_0^S \begin{pmatrix} -\sin(2\phi - 2\theta) & \cos(2\phi - 2\theta) \\ \cos(2\phi - 2\theta) & \sin(2\phi - 2\theta) \end{pmatrix}$$

, where $\sigma^K$ and $\sigma^{K'}$ are the conductivity contributed from different valleys and $\sigma_0^S = ne^2\tau \frac{(m_1 - m_2)}{m_1 m_2} \sin 2\phi$ is a parameter depending on the materials with carrier density $n$ and relaxation time $\tau$ (more details in Sec. IC of SI). The most remarkable property is that there will always be a spin current with a constant magnitude $|\sigma_0^S E_\theta|$ independent of the electric field direction. In details, the direction of spin current indeed depends on

6 / 20

the electric field direction $\theta$, and the most interesting case is $\theta = \phi$ or $\theta = \phi \pm \pi$ (i.e. electric field is in the mirror plane, see Fig. 1d). The charge current will be along the electric field direction, while the spin current will be perpendicular to the charge current, similar to the spin Hall effect[33], i.e. a spin current appears in the direction perpendicular to the charge current[30]. Different from in the spin Hall effect from strong SOC or noncollinear magnetic order, the transverse spin current in *C*-paired SVL originates from the anisotropic responses of valleys to the external electric field due to different effective mass along different directions, and the equivalent spin Hall angle is $S_\theta = \frac{(m_1 - m_2)\sin 2\phi}{m_1 + m_2 - (m_1 - m_2)\cos 2\phi}$, which is determined by the effective mass anisotropy and thus can be very large. At last, it is worth emphasizing that due to SVL and the absence of SOC, the depolarization time of the spin current can be very long. Thus, an AFM material of *C*-paired SVL can serve as an ideal device to generate pure spin current, which is highly desirable in AFM spintronics[27–30].

**C-paired SVL in monolayer V$_2$Se$_2$O.** Based on symmetry analysis and first-principles calculations, we find that the proposed *C*-paired SVL can be realized in monolayer V$_2$Se$_2$O, which we predict to be a 2D AFM material based on the crystal field theory, extensive first-principles calculations, the general Goodenough-Kanamori-Anderson rules[43–45], and model Hamiltonian analysis.

The existence of *C*-paired SVL in monolayer V$_2$Se$_2$O is rooted in its crystal symmetries. As shown in Fig. 2a, monolayer V$_2$Se$_2$O has a tetragonal structure consisting of three atomic layers with two V atoms and one O atom in the middle layer sandwiched by two Se-atom layers. Each V atom is surrounded by four Se and two O atoms as a distorted octahedron, possessing a local magnetic moment around 2 $\mu_B$, and the two V-atom sublattices form the 2D Néel AFM order (Fig. 2b). Moreover, these two sublattices are related by a mirror symmetry with respect to Se-O-Se plane (Fig. 2a) but *cannot* be transformed to each other by any translation operation, which, together with the AFM order, frees spin DOF and hence enables the *C*-paired SVL even without SOC.

The detailed band structures are calculated by DFT+U methods and verified by hybrid functional methods (see Methods and Fig. S15-16), and the results are shown in

Fig. 2c. There are two valleys at X and Y points related by the mirror symmetry (Fig. 2d). States around X and Y points are mainly from $V_1$ and $V_2$ atom respectively (Fig. S17-18) and indeed have opposite spins due to the Néel AFM order, which leads to the *C*-paired SVL in the absence of SOC. Even considering SOC, the form of SVL is unchanged, as enforced by the mirror symmetry (more details in Fig. S19).

The bulk $V_2Se_2O$ has VdW layered crystal structure and has been synthesized in experiments recently[46], and the monolayer $V_2Se_2O$ is thermal-dynamically stable (Fig. S23). Moreover, the cleavage energy of the bulk and in-plane stiffness of monolayer $V_2Se_2O$ is calculated to be comparable to graphite and graphene, which indicates that monolayer $V_2Se_2O$ can be easily exfoliated from the bulk (Fig. S23). Very recently, bulk $V_2Te_2O$ has also been synthesized[47]. Although they have the same crystal structure, their electronic and magnetic properties are completely different. $V_2Se_2O$ is insulating and has large local magnetic moments[46], while $V_2Te_2O$ is metallic with no local magnetic moments[47]. Based on our systematic calculations and analysis, these differences can be attributed to the different crystal field effect, which is stronger in $V_2Se_2O$ due to the larger electronic negativity of Se and shorter V-Se bonds. Moreover, the large crystal field effect in $V_2Se_2O$ also leads to stronger exchange interactions between local magnetic moments (~85 meV), which forces local magnetic moments to form the in-plane AFM alignment as shown in Fig. 2. With such an in-plane AFM alignment, bulk $V_2Se_2O$ is a semiconductor independent of the out-of-plane magnetic orders (Fig. S6-7), while $V_2Te_2O$ is always metallic even we assume it has the same magnetic structure (Fig. S9). Although the strong in-plane AFM ordering, bulk $V_2Se_2O$ only exhibit paramagnetic behavior[46], which is due to the small interlayer coupling (~0.11 meV) and in-plane magnetic anisotropy (~0.05 meV). All these theoretical results are consistent with and explain well experimental observations. Thus, we can conclude that monolayer $V_2Se_2O$ can have the AFM order even bulk $V_2Se_2O$ is paramagnetic, as the similar scenario for $VSe_2$, whose monolayer is strong room-temperature ferromagnetic while the bulk is paramagnetic[19]. In addition, due to very strong in-plane AFM interaction, one can use an external magnetic field to further stabilize the AFM order if the magnetic anisotropy turns out to be too small as for monolayer NiPS$_3$[24]. More details about the AFM order in monolayer V$_2$Se$_2$O can be found in Sec. II and Sec. III of SI.

Besides the conventional methods[22,48], one can also use spin-polarized scanning tunneling microscope/spectroscopy (STM/STS) to measure the magnetic structure directly in real space to the AFM order in monolayer V$_2$Se$_2$O. Moreover, the SVL offers another more convenient approach of simply using quasi-particle interference (QPI) as implemented in an ordinary STM. For an ordinary semiconductor, both the inter-valley scattering ($q_1$ and $q_2$) and intra-valley scattering ($q_3$ and $q_4$) can happen, and QPI patterns will exist around both the center and the corner like Fig. 2e. However, due to the $C$-paired SVL, the opposite spin of different valleys will strongly suppress the inter-valley scattering ($q_1$ and $q_2$), and QPI patterns will be only around the center but *not* the corner like Fig. 2f, i.e. the AFM order here can be easily confirmed by the absence of scatter patterns around the corner in QPI image. In addition to STM, one can use angle-resolved photoemission spectroscopy (ARPES) to measure the band structure and even use spin-resolved ARPES to directly measure the $C$-paired SVL.

**Piezomagnetism in monolayer V$_2$Se$_2$O.** As proposed above, the $C$-paired SVL can easily generate a valley polarization using a strain and may realize a giant piezomagnetism. As shown in Fig. 3a and S20, a compressive/tensile strain along $x$ direction makes the energy of $X$ valley $E(X)$ lower/higher than that of $Y$ valley $E(Y)$, and the valley polarization $P = E(X) - E(Y)$ depends on the strain monotonously (Fig. 3b). Meanwhile, the strain along $y$ direction generates exactly the opposite valley polarization since two valleys are related by a diagonal mirror symmetry (Fig. 3b).

To determine the magnetization upon finite doping accurately, we carefully tested the convergence of k grids using various methods (more details in Sec. IIIG). Without loss of generality, we focus on the hole doping. As shown in Fig. 3c, the magnetization increases with both carrier density and strain, and the magnetization direction is indeed opposite for tensile and compressive strains. The typical quantitative dependence is shown in Fig. 3d. Same as the general theory, the magnetization linearly depends on the strain in small strain region, and eventually saturates for very large strains, where all the carriers are polarized and hence the magnetization is always equal to the number of holes per unit cell. The piezomagnetic coefficient between magnetization and strain has some dependence on the carrier density, as the effective mass is no longer a constant when considering the high order effect of strain (Fig. S21). Similar results but with opposite magnetization direction happen when the strain is along $y$ direction since C-paired SVL here is enabled by the diagonal mirror symmetry (Fig. S22). Compared with piezomagnetism in noncollinear magnetic materials[41], piezomagnetism here is much larger, and the magnetization direction can be tuned using various methods since the AFM order is collinear and magnetic anisotropy energy is very small ($\sim 10^{-2}$ meV). So far, monolayer $V_2Se_2O$ is the first 2D material, which can realize the piezomagnetism.

**Noncollinear spin current generation in monolayer $V_2Se_2O$.** As shown in Fig. 2d, valleys of monolayer $V_2Se_2O$ are highly anisotropic, which suggests a giant noncollinear spin current generation as analyzed above. Without SOC, spin is a good quantum number and the conductivity of spin-up and spin-down electrons ($\sigma^{\uparrow,\downarrow}$) can be well defined and calculated. As shown in Fig. 4a, for electric field along $x$ direction, only the longitudinal current exists, and $\sigma_0^{\uparrow}$ is always smaller than $\sigma_0^{\downarrow}$ for the Fermi level in $[-2\text{eV}, 2\text{eV}]$, and the spin current is also along $x$ direction with the spin conductivity $\sigma_0^S = \sigma_0^{\uparrow} - \sigma_0^{\downarrow}$. To study the angle dependency of the spin current quantitatively, we choose Fermi level to be -0.2eV. As shown in Fig. 4b, both the longitudinal and transverse conductivity $\sigma_{L,T}^{\uparrow,\downarrow}$ oscillate with the electric field direction $\theta$ with a period of $\pi$, and can be well characterized by $\sigma_L^{\uparrow,\downarrow}(\theta) = \sigma_0 \mp \sigma_0^S \cos 2\theta$ and $\sigma_T^{\uparrow,\downarrow}(\theta) = \mp \sigma_0^S \sin 2\theta$. The longitudinal and transverse spin conductivity can be easily obtained as $\sigma_L^S(\theta) = \sigma_L^{\uparrow} - \sigma_L^{\downarrow} = -2\sigma_0^S \cos 2\theta$ and $\sigma_T^S(\theta) = \sigma_T^{\uparrow} - \sigma_T^{\downarrow} = 2\sigma_0^S \sin 2\theta$ (Fig. 4c). Same as the general theory, the noncollinear spin current is a constant independent of the electric field direction. As shown in the Fig. 4d, directions of both charge current and spin current depends on the electric field direction. Although there are transverse charge currents from individual valleys, while they cancel with each other and hence the charge current is always along the electric field direction $\theta$. However, the spin currents from two valleys cannot cancel, resulting in a net spin current in the direction $\pi - \theta$ (Fig. 4d), linear in the electric field strength. Moreover, when $\theta = \pi/4$ and $\theta =$ $3\pi/4$, the spin current will be perpendicular to the charge current (and electric field), and the equivalent spin Hall angle is 0.7, much larger than the reported values due to SOC[49,50]. The spin splitting here (> 1 eV as shown in Fig. 3c and S16c) is also much larger than in the *T*-paired SVL. Together with the absence of or very weak SOC, a long depolarization time of spin current can be very long, which is critical for the spintronics applications.

Besides in 2D AFM materials, *C*-paired SVL can generally exist in 3D AFM materials. Based on symmetry analysis and first-principles calculations, we found C-paired SVL can also exist in the following experimentally verified AFM materials, $NaOsO_3$, $LaMnO_3$, $LaCrO_3$, $TbFeO_3$, MnTe, $RuO_2$, $MnF_2$, $FeF_2$, $CoF_2$ and $NiF_2$. Due to the complexity of 3D magnetic space groups and band structures, we will put the detailed studies of these materials in another work. In addition, a complete search of *C*-paired materials could be done using the high-throughput method in the future work.

In conclusions, we proposed a new type of SVL, *C*-paired SVL, which has many advantages like giant piezomagnetism and noncollinear spin current. Via first-principles calculations, we predicted that monolayer $V_2Se_2O$, whose bulk compound has been synthesized recently, can host the *C*-paired SVL, which is also the first 2D piezomagnetic material under doping. In view of experimental studies, the angle-dependent spin current can be measured with standard schemes, the 2D Néel magnetic structure in real space can be visualized by spin-polarized STM/STS, and the *C*-paired SVL in momentum space can be directly detected using spin-resolved ARPES or QPI as implemented in STM. External strain can be induced by depositing $V_2Se_2O$ on a substrate with moderate lattice mismatch such as (001) surface of MgO (4.21 Å).

**Methods**

The calculations are performed in the framework of density functional theory as implemented in Vienna *ab initio* simulation package (VASP)[51]. The projector-augmented wave potential is adopted with the plane-wave energy cutoff set at 600 eV. The exchange-correlation functional of the Perdew-Burke-Ernzerhof (PBE) type has been used for both structural relaxations and self-consistent electronic calculations[51,52]. The GGA+U method is employed to treat the strong correlations of the V 3$d$ orbitals[53], where the value of the Hubbard $U$ is taken as 5.1eV, and the Hund's rule coupling $J$ is taken as 0.8 eV. These two interaction parameters are determined by fitting to the band structure obtained from the hybrid functional calculations[54]. The Brillouin zone (BZ) is sampled by a 9×9×1 gamma-centered Monkhorst-Pack mesh, and the convergence criteria for the electronic iteration is set to $10^{-6}$ eV. Denser k mesh has been very carefully tested to give converged results in the calculations of piezomagnetism and conductivity.

**Acknowledgements** We thank for the helpful discussions with Vic K. T. Law, Chen Fang, Hua Jiang and Xi Dai. This work is supported by the Hong Kong Research Grants Council (ECS26302118, N_HKUST626/18). We also knowledge the financial support from National Natural Science Foundation of China (11521404, 11634009, 11674222, 11674226, 11790313, 11574202, 11874256, U1632102, 11861161003 and 11874258), the National Key Research and Development Program of China (Grant Nos. 2016YFA0300403, 2016YFA0301003). This work is supported in part by the Key Research Program of the Chinese Academy of Sciences (XDPB08-2), the Strategic Priority Research Program of Chinese Academy of Sciences (XDB28000000), and the General Research Fund of Hong Kong RGC (HKU17303518). The computation is performed in the Tianhe2 cluster at the National Supercomputer Center in Guangzhou, China.

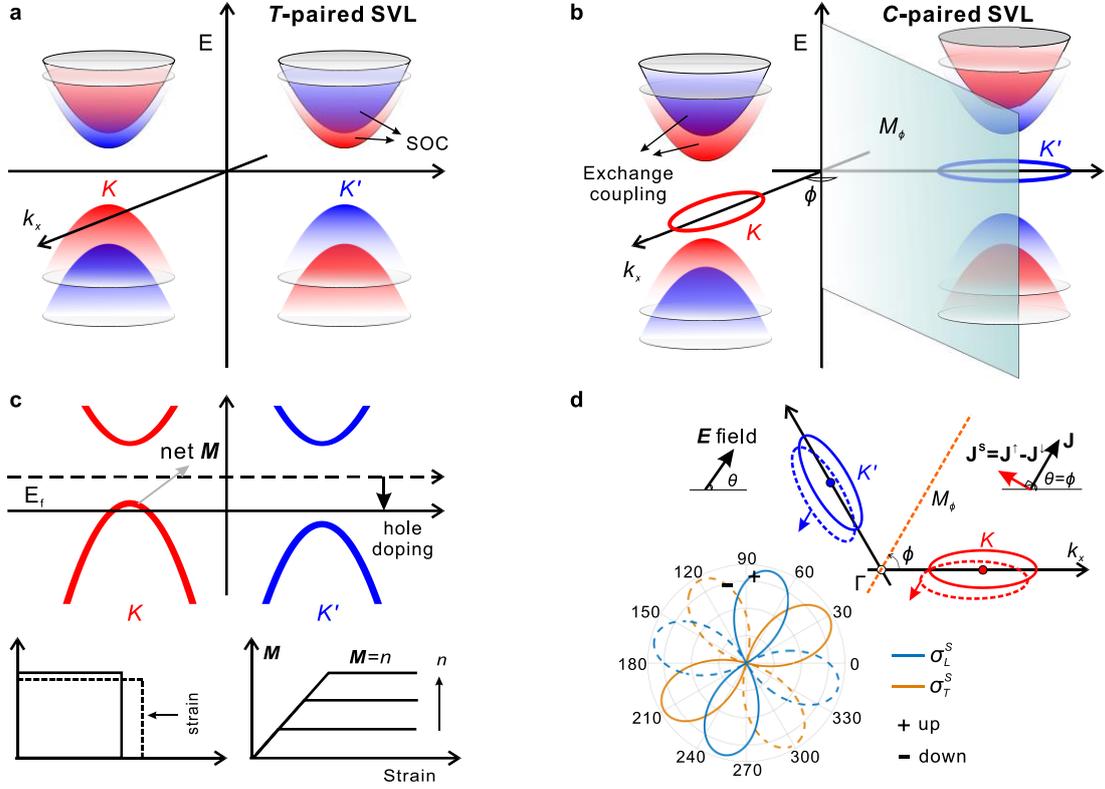

**Fig. 1 | Schematics of *C*-paired spin-valley locking (SVL) and its properties. a,** In *T*-paired SVL, the splitting between spin-up (red) and spin-down (blue) bands is from strong spin-orbit coupling (SOC) when inversion is broken, and the two valleys are related by the time reversal (*T*) symmetry. **b,** However, for *C*-paired SVL, the spin splitting is due to exchange couplings between itinerant electrons and local magnetic moments, and different valleys are related by a crystal symmetry (a mirror symmetry $M_\phi$ as an example). **c,** A strong valley polarization can be easily induced by a strain, and the piezomagnetism will simultaneously happen upon finite doping. *M* first increase linearly with the strain and finally gets saturated for large strain with *M* equal to *n*. **d,** Contribution to the charge current from different valleys are in general different, which gives rise to a non-zero spin current $J^S = J_K - J_{K'}$. One typical result ($\phi = \frac{\pi}{6}$) is in the bottom panel. Both longitudinal and transverse spin currents depend on the

17 / 20

electric field $E$ direction $\theta$ with a period of $\pi$. When $\theta = \phi$ or $\phi + \pi$, where $E$ does not break the mirror symmetry, the longitudinal charge currents from different valleys are the same, while transverse charge currents are exactly opposite, which result in pure spin currents perpendicular to the charge currents, i.e. spin Hall effect.

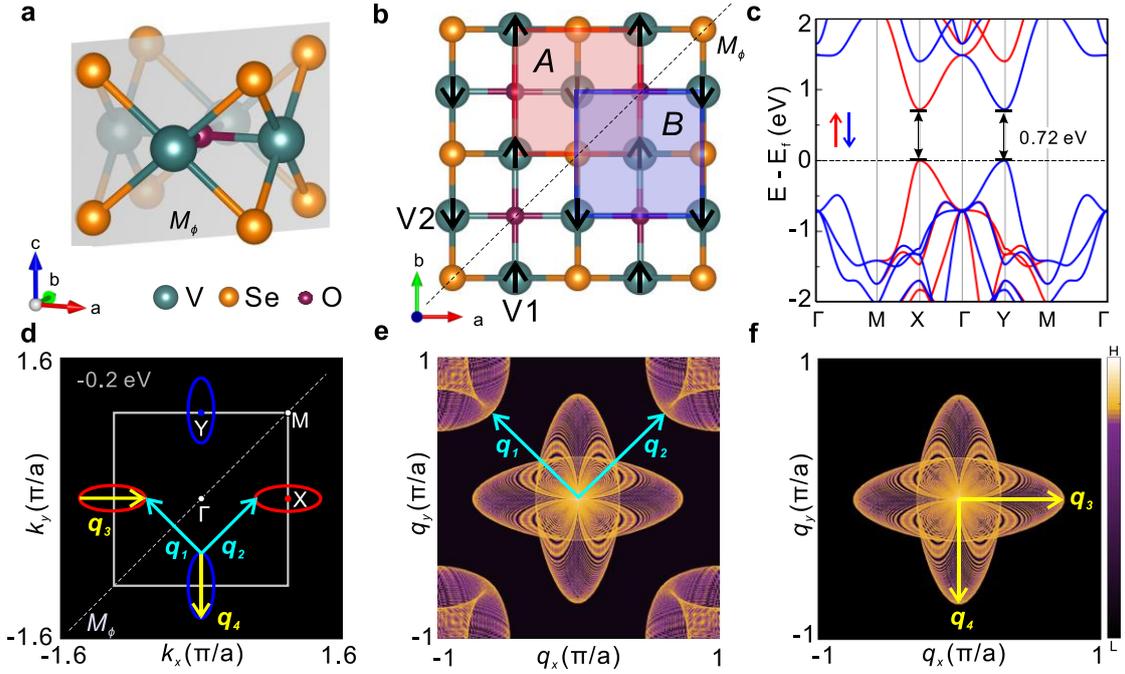

**Fig. 2 | $C$-paired SVL in monolayer $V_2Se_2O$. a,** Two sublattice A and B in monolayer $V_2Se_2O$ is related by a diagonal mirror symmetry $M_\phi$. **b,** 2D Néel anti-ferromagnetic (AFM) order with local magnetic moments located at V1 and V2 atoms with polarization labelled by black arrows. **c,** Monolayer $V_2Se_2O$ is a semiconductor with gap around 0.7 eV. Red (blue) color is for spin up (down). **d,** The energy contour taken at 0.2 eV below the valance band maximum with possible typical inter-valley and intra-valley scattering vectors $q_{1,2}$ and $q_{3,4}$. (**e, f**) Calculated quasi-particle interference (QPI) patterns with and without considering the spin flipping. Due to SVL and weak SOC, inter-valley scattering is strongly suppressed. The existence of AFM order and $C$-paired SVL can be easily verified by the absence of scattering patterns around the corner in QPI image.

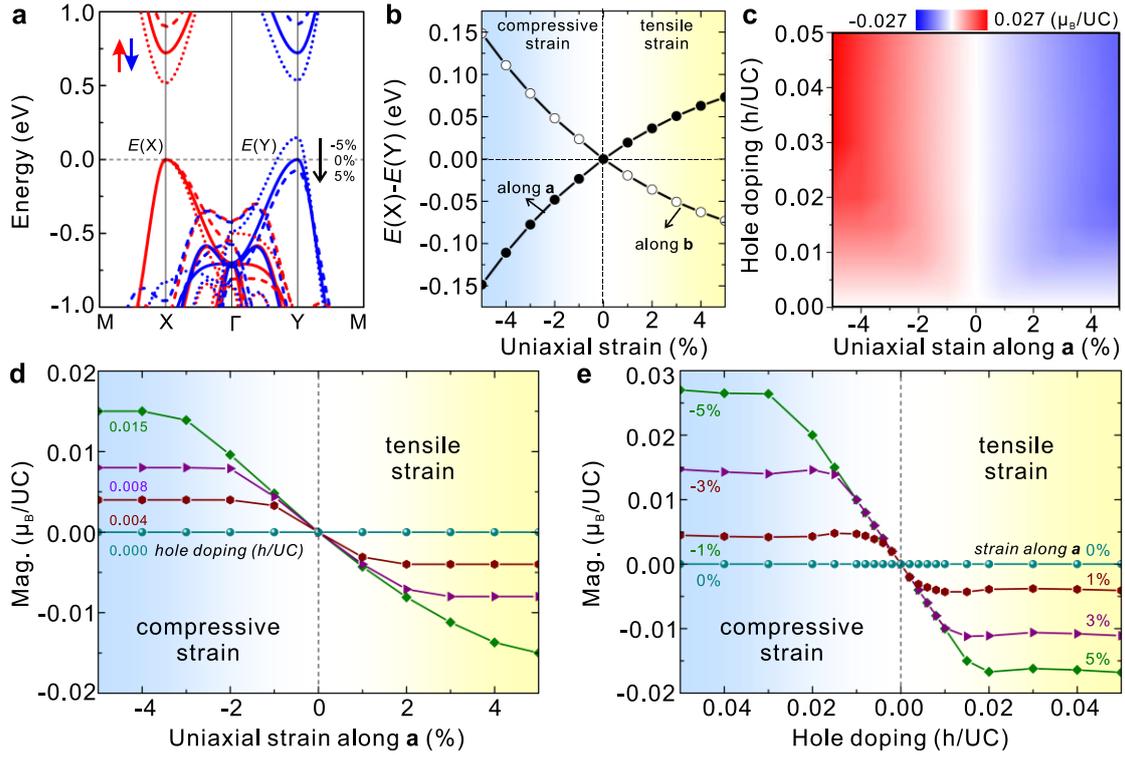

**Fig. 3 | Strain induced valley polarization and piezomagnetism in hole doped monolayer V₂Se₂O. a,** Band structure evolution under different uniaxial strains along $x$ direction (-5%, 0% and -5%). Compared to the energy $E(X)$ at $X$ valley, the energy $E(Y)$ at $Y$ valley monotanesouly shifts down. **b,** Strain induced valley polarization, which defined as the energy difference between two valleys $P = E(X) - E(Y)$. **c,** Diagram of strained induce net magnetization for different hole densities. **d,** For a given hole density, the net magnetization increases linearly with strain but has opposite direction for compressive and tensile strains in the region of small strains. When strain is large enough, all the carriers are polarized, and the net magnetization saturates. **e,** For any given strain, the magnetization is always equal to the number of holes per unit cell when the doping is light. However, for heavy doping, the magnetization is almost a constant that depends on the strain.

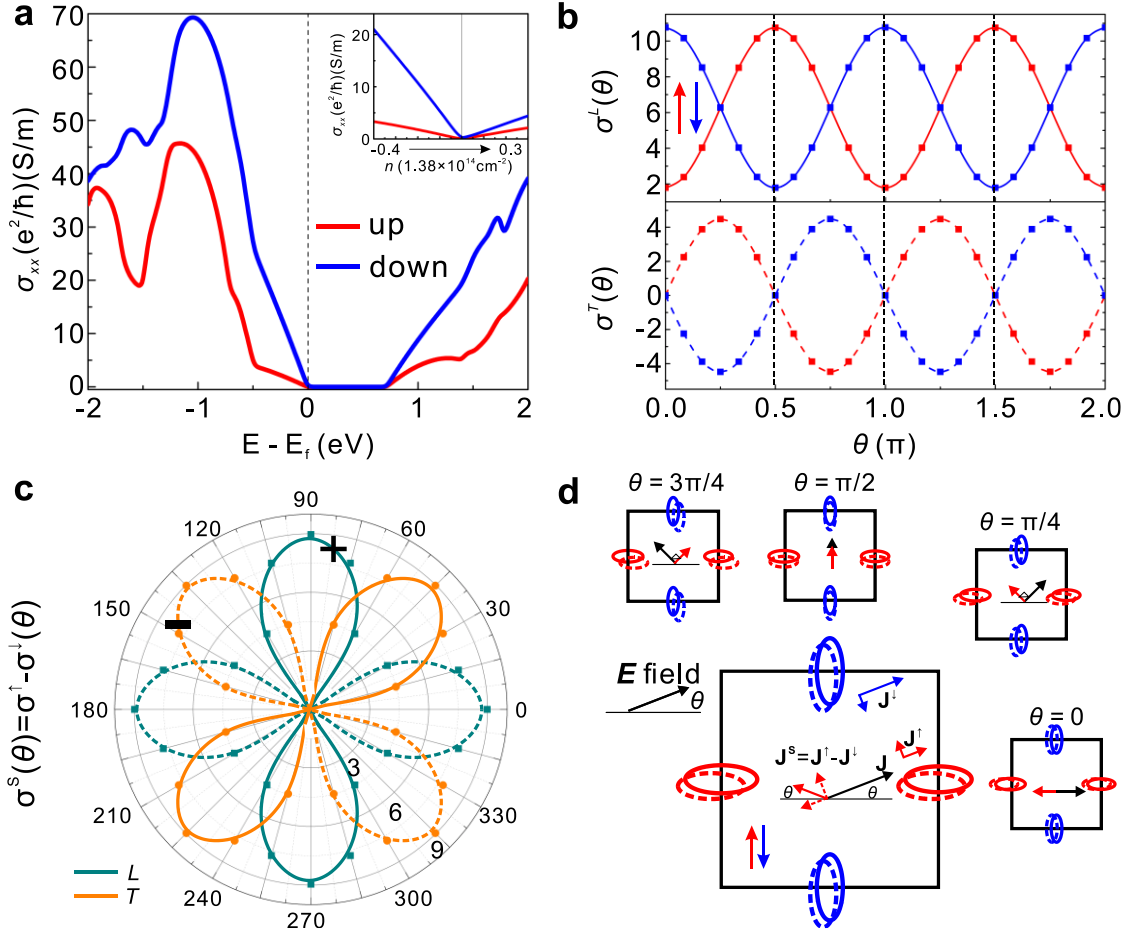

**Fig. 4 | Angle-dependent spin current and spin Hall effect in monolayer V$_2$Se$_2$O. a,** Spin-resolved charge conductivity $\sigma$ for electric field along $x$ direction. Inset is the dependency of $\sigma$ on the carrier density $n$. For light doping, $\sigma$ increases almost linearly with $n$. **b,** Angle-dependence of the longitudinal (*L*) and transverse (*T*) charge conductivity varying with the electric field direction $\theta$, taken at 0.2 eV below the valance band maximum. The transverse charge conductivity is always opposite for spin-up and spin-down electrons, and hence the transverse charge current is always zero and the charge current direction is always $\theta$. **c,** The corresponding angle-dependence of the longitudinal and transverse spin conductivity. **d,** Relation between directions of charge current *J* and spin current *J$^S$*. The spin current always flows along the direction $\pi - \theta$. In details, for $\theta = \pm\pi/2$, the spin current will be along the same direction of the charge current, while for $\theta = 0$ or $\pi$, the spin current flow along the opposite direction of the charge current. Moreover, for $\theta = \pm\pi/4$ or $\pm 3\pi/4$, there will be large spin Hall effect, where the spin current is perpendicular to the charge current.